\documentclass[aps,twocolumn,showpacs]{revtex4}%
\usepackage{amsmath}
\usepackage{graphicx}
\usepackage{amsfonts}
\usepackage{amssymb}%
\begin{document}
\title{Non-exponential relaxation and quantum tunnel splitting in molecular magnet
Fe$_{8}$}
\author{Zhi-De Chen$^{1,2}$ and Shun-Qing Shen$^{1}$}
\affiliation{$^{2}$Department of Physics and Institute of Modern Condensed Matter Physics,
Guangzhou University, Guangzhou 510405, China}
\date{July 24, 2002}

\begin{abstract}
Magnetic relaxation in molecular magnets under a sweeping field is studied by
taking into account local stray fields. It is found that the randomness of
local stray field leads to a distribution of the relaxation rate which
subsequently makes the relaxation deviate from the exponential law as
predicted by the Landau-Zener model such that the Landau-Zener method needs to
be revised to deduce an exact tunneling splitting. The tunneling splitting and
distribution width of local stray fields are derived from the experimental
data for molecular magnets Fe$_{8}$.

\end{abstract}
\maketitle

Magnetic relaxation by quantum tunneling in high spin molecular magnets has
become an attractive field of research in recent years[1-15]. One of the
well-studied systems is octanuclear iron(III) oxo-hydroxo clusters Fe$_{8}$,
which has a well-defined temperature independent region below
0.36K.\cite{Sangre97,Wernsdor99, Wernsdor99b,Wernsdor00,Wernsdor01}
Theoretically the key to understand magnetic relaxation by the quantum
tunneling is the tunnel splitting. The tunnel splitting and the magnetization
relaxation are connected via the Landau-Zener
model.\cite{Barbara99,LZ,Miyashita95} Since the Landau-Zener transition rate
is explicitly related to the tunnel splitting, the measurement of the change
of the magnetization after one sweep over the resonant point can give the
tunnel splitting\cite{Miyashita95}. Usually the tunnel splitting is very small
(like the ground state tunneling of Fe$_{8}$), and the transition rate due to
tunneling is also very low. One sweep cannot lead to an observable
magnetization relaxation. In this case, the tunnel splitting is determined by
the Landau-Zener method: \cite{Wernsdor99,Wernsdor00,Wernsdor01} multi-sweeps
are done and the tunnel splitting is deduced from the experimental data in a
short time region. Up to now, the Landau-Zener
method\cite{Wernsdor99,Wernsdor00,Wernsdor01} has served as a basic tool to
study quantum tunneling in molecular magnets and many other interesting
phenomena including the oscillation of tunnel splitting with respect to the
field along the hard axis and the parity effect for odd and even
resonance.\cite{Garg93, Wernsdor99,Chen02} The tunnel splitting from the
Landau-Zener method is found to be sweeping-rate-independent and agrees with
the result found by using a square-root decay
method.\cite{Wernsdor99,Wernsdor00,Wernsdor01} However, there is still a
puzzle that magnetization relaxation under a sweeping field shows a clear
deviation from the exponential behavior as predicted by the Landau-Zener
model.\cite{Wernsdor00,Wernsdor01} Such a consequence may lead to a question
how the tunnel splitting is deduced from the experimental data in short time
region exactly. This is the main motivation of the present paper.

In the present paper, we start with the biaxial spin model with a local stray
field to study the magnetization relaxation behaviors. It is shown that the
\textquotedblleft uncompensated\textquotedblright\ transverse component of the
local stray field leads to a distribution of relaxation rates which makes the
relaxation in molecular magnets follow a different mechanism as in some
complex systems like spin-glass.\cite{Palmer84,Erhart94} Our analysis shows
that the relaxation is determined by two independent quantities: the tunnel
splitting and the distribution width of the local stray field. Although the
magnetization relaxation deviates the exponential law we can derive the two
quantities from experimental data of molecular magnets Fe$_{8}$ successfully.

The biaxial spin model for the molecular magnets Fe$_{8}$ with a local stray
field is written as\cite{Chund88,Chen02, Garg93}%

\begin{equation}
H=K_{1}S_{z}^{2}+K_{2}S_{y}^{2}-g\mu_{B}\mathbf{S\cdot(B+h),}%
\end{equation}
where $K_{1}>K_{2}>0$ , and $\mathbf{B}$ is the applied magnetic field.
$\mathbf{h}$ is the local stray field which may originate from the
interactions between the giant spin and the environmental spins (including
other giant spins or nuclear spins). To simplify the problem we assume that
$\mathbf{h}$ has a Gaussian distribution with an equal distribution width in
all directions
\begin{equation}
P(\mathbf{h})=\frac{1}{(2\pi\sigma^{2})^{3/2}}\exp\left[  -(\mathbf{h}%
-\mathbf{h}_{0})^{2}/2\sigma^{2}\right]  .
\end{equation}
Since our main interest is the magnetization relaxation under a sweeping
field, the external magnetic field is taken to be$\ \mathbf{B}=\{B_{x},0,0\}$:
$B_{x}=n\Delta B\pm ct$ where $n$ is integer, $\Delta B$ is the field interval
between neighboring resonant tunneling and $c=dB_{x}/dt$. In the following
calculation, we take $K_{1}=0.321$K, $K_{2}=0.229$K for the molecular magnets
Fe$_{8}.$\cite{Wernsdor99} When the field along the easy axis is sweeping over
the resonant field $n\Delta B$, the biased local stray field $h_{x}$ will be
compensated by the sweeping field which brings spins into the resonant
tunneling and leads to a continuous relaxation. If we omit the transverse
component of the local stray field, then all the spins will have the same
tunneling rate inside the resonant window. The resulted relaxation is the
simple exponential decay according to the Landau-Zener
model\cite{Wernsdor00,Wernsdor01}
\begin{equation}
M(t)=M_{0}e^{-\Gamma t}%
\end{equation}
where $\Gamma=kP_{LZ}c/A$ and $P_{LZ}$ is the Landau-Zener transition rate,
$P_{LZ}=1-\exp\left(  -\pi\Delta_{n}^{2}/\nu_{n}c\right)  ,$ $\Delta_{n}$ is
the tunnel splitting, $\nu_{n}=2g\mu_{B}\hbar(2s-n),$ $k=2$ for $n=0$ and
$k=1$ for $n=1,2,3,\cdot\cdot\cdot,$ and $A$ is the amplitude of the AC field
used in the experiment. \ In the low transition rate limit, i.e., $P_{LZ}%
\ll1,$ which holds for Fe$_{8}$ system\cite{Wernsdor00,Wernsdor01}, the above
equation leads to
\[
\ln\frac{M(t)}{M_{0}}=-\frac{k\pi}{\nu_{n}A}\Delta_{n}^{2}t+\frac{1}{2}\left(
\frac{\pi\Delta_{n}}{\gamma_{n}c}\right)  ^{2}\frac{kc}{A}t+\cdots
\]
In the Taylor series expansion, one has
\begin{equation}
\frac{M_{0}-M(t)}{M_{0}}=\frac{k\pi t}{\nu_{n}A}\Delta_{n}^{2}-\frac{1}%
{2}\left[  \left(  \frac{k\pi t}{\nu_{n}A}\right)  ^{2}+\frac{\pi^{2}%
kt}{\gamma_{n}^{2}cA}\right]  \Delta_{n}^{4}+\cdot\cdot\cdot. \label{dm1}%
\end{equation}
\ In a short time region, we can keep the first term on the right side of
Eq.(\ref{dm1}), and then the tunnel splitting $\Delta_{n}$ can be deduced from
the magnetization relaxation of the molecular magnets.

However, the tunnel splitting strongly depends on the strength of the
transverse local stray field. In other words, the local stray field will leads
to a distribution of the tunneling splitting, and furthermore a distribution
of the relaxation rates. Consequently, the resulted relaxation is modified as
\begin{equation}
M(t)=M_{0}\int d\mathbf{h}P(\mathbf{h})e^{-\Gamma(\mathbf{h})t}, \label{mt1}%
\end{equation}
where
\begin{equation}
\Gamma(\mathbf{h})=[1-\exp\left(  -\pi\Delta_{n}^{2}(\mathbf{h})/\nu
_{n}c\right)  ]kc/A. \label{gamma}%
\end{equation}
The tunnel splitting $\Delta_{n}^{2}(\mathbf{h})$ can be calculated by the
instanton method,\cite{Chen02,Chen}
\begin{equation}
\Delta_{n}^{2}(\mathbf{h})\simeq\gamma_{\pm}(\mathbf{h})\Delta_{n0}^{2}
\label{splitting}%
\end{equation}
where the renormalized factor is caused by the local stray field, $\gamma
_{\pm}(\mathbf{h})=[\cosh(2qh_{y})\pm\cos(2d_{n}h_{z})]/2$ and $\pm$ stands
for the even and odd resonant tunneling.
\begin{equation}
d_{n}=\frac{g\mu_{B}}{2K_{1}}\int_{0}^{\pi}\frac{d\phi}{1-\frac{K_{2}}{K_{1}%
}\sin^{2}\phi-\frac{g\mu_{B}n\Delta B}{2sK_{2}}\cos\phi},
\end{equation}
$q=g\mu_{B}\pi/2(K_{1}K_{2}-K_{2}^{2})^{1/2},$ and $\Delta_{n0}$ is
independent of the transverse field. \ In the low transition rate
limit,\cite{Wernsdor00,Wernsdor01} we have
\begin{equation}
M(t)\simeq M_{0}\int d\mathbf{h}P(\mathbf{h})\exp\left\{  -\gamma_{\pm
}(\mathbf{h})\Gamma_{0}t\right\}  , \label{m(t)}%
\end{equation}
where $\Gamma_{0}=k\pi\Delta_{n0}^{2}/(\nu_{n}A)$.%

%TCIMACRO{\FRAME{ftbpFU}{3.1211in}{2.3099in}{0pt}{\Qcb{Short time relaxation
%behavior for both odd and even resonant tunneling with $\sigma=0.08$T and
%$h_{0}=\sigma/4.$}}{}{fig1.eps}{\special{ language "Scientific Word";
%type "GRAPHIC";  display "USEDEF";  valid_file "F";  width 3.1211in;
%height 2.3099in;  depth 0pt;  original-width 3.9972in;
%original-height 3.1583in;  cropleft "0";  croptop "1";  cropright "1";
%cropbottom "0";  filename 'fig1.eps';file-properties "XNPEU";}}}%
%BeginExpansion
\begin{figure}
[ptb]
\begin{center}
\includegraphics[
height=2.3099in,
width=3.1211in
]%
{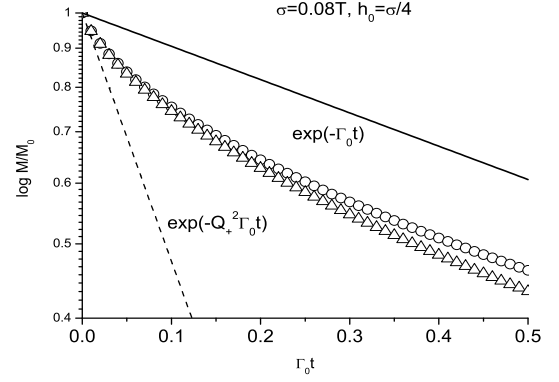}%
\caption{Short time relaxation behavior for both odd and even resonant
tunneling with $\sigma=0.08$T and $h_{0}=\sigma/4.$}%
\end{center}
\end{figure}
%EndExpansion

The above equation shows that, in the low transition rate limit, the magnetic
relaxation is sweeping-rate-independent, provided that the sweeping rate is
large enough, namely, for $dB_{x}/dt\gtrsim1.0$ mT/s.\cite{Wernsdor00,Chen} In
the absence of the transverse local stray field, i.e., $P(\mathbf{h}%
)=\delta(\mathbf{h}),$ we have $M(t)\simeq M_{0}e^{-\Gamma_{0}t}$ for an even
resonant tunneling and $M(t)=M_{0}$ for an odd resonant tunneling as expected.
The effect of the transverse local stray field can be observed by doing the
integration in Eq.(\ref{m(t)}). Numerical results \cite{note1} as shown in
Fig.1 show that the magnetization relaxation deviates apparently from the
exponential decay as the distribution width becomes larger than $0.03$T, which
is in a qualitative agreement with the experimental measurement. The
non-exponential decay indicates the Landau-Zener model cannot be applied to
measure the tunneling splitting explicitly. In fact, for a system with a
distribution of relaxation time, the resulted relaxation can be a large
variety of shapes of decay.\cite{Palmer84, Erhart94} The above analysis shows
that we cannot deduce the tunnel splitting according to Eq.(\ref{dm1}) when
the relaxation deviates strongly from the exponential law. On the other hand,
consider the local stray fields one obtains
\begin{equation}
\frac{M_{0}-M(t)}{M_{0}}=\frac{k\pi t}{\nu_{n}A}\langle\Delta_{n}^{2}%
\rangle-\frac{1}{2}\left[  \left(  \frac{k\pi t}{\nu_{n}A}\right)  ^{2}%
+\frac{\pi^{2}kt}{\gamma_{n}^{2}cA}\right]  \langle\Delta_{n}^{4}\rangle
+\cdot\cdot\cdot\label{decay}%
\end{equation}
where $\langle X\rangle=\int d\mathbf{h}P(\mathbf{h})X$. Comparing with
Eq.(\ref{dm1}), one can see that the tunnel splitting determined from
$[M(t)-M_{0}]/M_{0}$ under such an approximation is $\sqrt{\langle\Delta
_{n}^{2}\rangle}$ instead of $\Delta_{n}$. From Eq.(\ref{splitting}), one has
\begin{equation}
\sqrt{\langle\Delta_{n}^{2}\rangle}=Q_{\pm}\Delta_{n0},\qquad\label{d1}%
\end{equation}
where the averaging renormalization factor due to the local stray fields is
\begin{equation}
Q_{\pm}=\frac{1}{\sqrt{2}}\left[  e^{2q^{2}\sigma^{2}}\cosh(2qh_{0})\pm
e^{-2d_{n}^{2}\sigma}\cos(d_{n}h_{0})\right]  ^{1/2}.
\end{equation}
Dependence of $Q_{\pm}$ on\ the distribution width $\sigma$ is plotted in
Fig.2. It is seen that $Q_{\pm}$ increase rapidly with increasing $\sigma$ and
become coincident as $\sigma>0.08$T. For an even resonance, $Q_{+}\geqslant1$,
and $\sqrt{\langle\Delta_{n}^{2}\rangle}$ is always larger than the exact
tunnel splitting $\Delta_{n0}$, while for an odd resonance, the tunnel
splitting\ $\sqrt{\langle\Delta_{n}^{2}\rangle}$ is not always quenched once
the local stray field appears.%
%TCIMACRO{\FRAME{ftbpFU}{3.0813in}{2.3557in}{0pt}{\Qcb{Illustration of the
%dependence of $Q_{\pm}$ on the distribution width $\sigma$. The dashed and
%solid lines represent for Q$_{+}$ a Q$_{-}$, respectively.}}{}{fig2.eps}%
%{\special{ language "Scientific Word";  type "GRAPHIC";  display "USEDEF";
%valid_file "F";  width 3.0813in;  height 2.3557in;  depth 0pt;
%original-width 4.0802in;  original-height 3.1176in;  cropleft "0";
%filename 'fig2.eps';file-properties "XNPEU";}}}%
%BeginExpansion
\begin{figure}
[ptb]
\begin{center}
\includegraphics[
height=2.3557in,
width=3.0813in
]%
{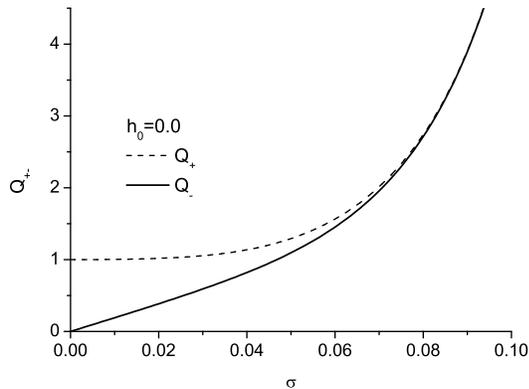}%
\caption{Illustration of the dependence of $Q_{\pm}$ on the distribution width
$\sigma$. The dashed and solid lines represent for Q$_{+}$ a Q$_{-}$,
respectively.}%
\end{center}
\end{figure}
%EndExpansion

Now we are ready to derive the tunneling splitting from the experimental data
of Fe$_{8}.$ The experimental data of the ground state tunneling in Fe$_{8}$
were provided by Wernsdorfer.\cite{Wernsdor00,Wernsdor01} To fit the
experimental data we choose $\Gamma_{0}=6.5\times10^{-4}$/sec and
$\sigma=0.05$T. Both theoretical and experimental results are plotted in
Fig.3. One can find that our analytic result fits the experimental curve quite
well. The estimated value $\sigma=0.05T$ is consistent with the linewidth of
the resonance for Fe$_{8}.$\cite{Prokof98} However, the hole-digging
method\cite{Wernsdor99b,Wernsdor01} gives a value 0.03T, which is smaller than
what we estimated. The tunnel splitting from the chosen $\Gamma_{0}$ is
$\Delta_{0}=7.85\times10^{-8}$ K, and the renormalization factor is
$Q_{+}(\sigma=0.05T)\simeq1.292$. Therefore we have $\sqrt{\langle\Delta
_{0}^{2}\rangle}\simeq1.01\times10^{-7}$K, which is very closed to
$1.0\times10^{-7}$K in the conventional Landau-Zener model by Wernsdorfer et
al. \cite{Wernsdor00,Wernsdor01} So what measured in the Landau-Zener model
is$\sqrt{\langle\Delta_{0}^{2}\rangle}$ instead of $\Delta_{0}$ in the short
time limit. Clearly, $\Delta_{0}$ and $\sqrt{\langle\Delta_{0}^{2}\rangle}$
are two different concepts. For Fe$_{8}$, since the averaging renormalization
factor is as large as 1.292, we should consider the effect from the local
stray field. Our result shows that the magnetization relaxation is determined
by the tunnel splitting and the distribution width of the local stray field.
The later quantity leads to the relaxation deviating from the exponential
decay. On the other hand, after the local stray field is introduced, the main
modification to the kinetic equation of the relaxation is to replace
$\Delta_{n0}$ with $\sqrt{\langle\Delta_{n}^{2}\rangle}$ in the low transition
rate limit.
%TCIMACRO{\FRAME{ftbpFU}{3.0995in}{2.2978in}{0pt}{\Qcb{Relaxation curve of
%ground state tunneling in Fe$_{8}$ molecule using a sweeping field with
%$A=7.2495\times10^{-2}$ T and $dB_{x}/dt=1.4$mT/s.}}{}{fig3.eps}%
%{\special{ language "Scientific Word";  type "GRAPHIC";  display "USEDEF";
%valid_file "F";  width 3.0995in;  height 2.2978in;  depth 0pt;
%original-width 3.9124in;  original-height 3.1583in;  cropleft "0";
%filename 'fig3.eps';file-properties "XNPEU";}}}%
%BeginExpansion
\begin{figure}
[ptb]
\begin{center}
\includegraphics[
height=2.2978in,
width=3.0995in
]%
{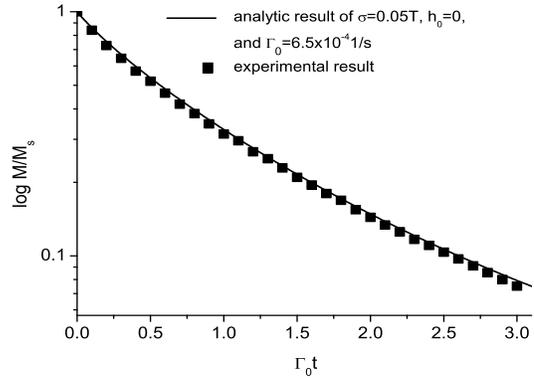}%
\caption{Relaxation curve of ground state tunneling in Fe$_{8}$ molecule using
a sweeping field with $A=7.2495\times10^{-2}$ T and $dB_{x}/dt=1.4$mT/s.}%
\end{center}
\end{figure}
%EndExpansion

%

%TCIMACRO{\FRAME{ftbpFU}{3.058in}{2.2234in}{0pt}{\Qcb{Distribution of ground
%state tunnel splitting for a Fe$_{8}$ model with higher order term
%$C(S_{+}^{4}+S_{-}^{4})$ in Hamiltonian under a local stray field with a
%Gaussian distribution. ($C=-2.95\times10^{-5}$K)}}{}{fig4.eps}%
%{\special{ language "Scientific Word";  type "GRAPHIC";  display "USEDEF";
%valid_file "F";  width 3.058in;  height 2.2234in;  depth 0pt;
%original-width 3.9669in;  original-height 3.1583in;  cropleft "0";
%filename 'fig4.eps';file-properties "XNPEU";}}}%
%BeginExpansion
\begin{figure}
[ptb]
\begin{center}
\includegraphics[
height=2.2234in,
width=3.058in
]%
{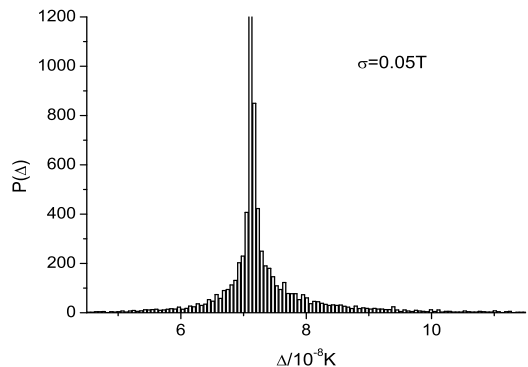}%
\caption{Distribution of ground state tunnel splitting for a Fe$_{8}$ model
with higher order term $C(S_{+}^{4}+S_{-}^{4})$ in Hamiltonian under a local
stray field with a Gaussian distribution. ($C=-2.95\times10^{-5}$K)}%
\end{center}
\end{figure}
%EndExpansion

It should be noted that the tunnel splitting from the experimental data is
sweeping-rate independent since the magnetic relaxation is independent of the
sweeping rate when $c>1.0$mT/sec. From Eq. (\ref{gamma}) we found that,
rigorously speaking, the magnetization relaxation depends on the sweeping rate
c via the expression for $\Gamma(\mathbf{h}).$ However, the larger c leads to
a smaller transition rate. When the low transition rate approximation becomes
valid for a sufficiently large c, the transition rate is independent of c.
This is consistent with the experimental measurement. Another problem is that
the above analysis is based on the adiabatic approximation. It is assumed that
all the spins have sufficient time to tunnel no matter how small the tunnel
splitting is. When the distribution width of the local stray field is taken
into account, the distribution of tunnel splitting becomes quite large.
Approximately there will be a cut-off tunnel splitting $\Delta_{nc}\simeq
\nu_{n}c/\pi$ such that the spins with the tunnel splitting lower than
$\Delta_{nc}$ will have no sufficient time to tunnel.\cite{Chund01,Mertes01}
To find out the distribution of the tunnel splitting due to the local stray
field, we have made Monte-Carlo simulation for Fe$_{8}$ systems with higher
order terms in the Hamiltonian. The resulted distribution of tunnel splitting
is plotted in Fig.4 for $\sigma=0.05$T. The tunnel splitting spreads over
about 2 to 3 order which is much narrower than that in Mn$_{12}$ molecules due
to dislocation.\cite{Chund01,Mertes01} As the local stray field from
dipolar-dipolar or hyperfine interaction in Mn$_{12}$ is stronger than that in
Fe$_{8}$ molecules and dislocation will lead to a wider distribution of the
tunnel splitting. The Mn$_{12}$ system will have a much wider distribution of
relaxation time than Fe$_{8}$ system. It implies that the magnetization
relaxation in Mn$_{12}$ system will deviate far away from the exponential decay.

In conclusion we have studied the effect of the local stray field on the
magnetic relaxation under a sweeping field in Fe$_{8}$ molecules. The
uncompensated transverse local stray field leads to a distribution of the
transition rate such that the relaxation deviates from the exponential law.
The interplay of the quantum tunneling and the distribution of the local stray
field determines the magnetization relaxation. Based on this picture we
proposed a revised scheme to deduce the tunnel splitting $\Delta_{n}$ from
experimental measurement instead of the conventional Landau Zener method. Our
conclusion can be generalized to other molecular magnets since the local stray
field due to the dipolar-dipolar or hyperfine interaction exists extensively.

The authors thank W. Wernsdorfor for his providing the experimental data in
Fig.3. The work was supported by a grant from the Research Grants Council of
the Hong Kong, China, and a CRCG\ grant of the University of Hong Kong.


\begin{thebibliography}{99}                                                                                               %


\bibitem {Gunther94}L. Gunther and B. Barbara, \textit{Quantum tunnelling of
Magnetization} --- QTM'94, (Kluwer, Dordrecht, 1995).

\bibitem {Sangre97}C. Sangregorio, T. Ohm, C. Paulsen, R. Sessoli, and D.
Gatteschi, Phys. Rev. Lett. \textbf{78}, 4645 (1998).

\bibitem {Wernsdor99}W. Wernsdorfer and R. Sessoli, Science \textbf{284}, 133 (1999).

\bibitem {Wernsdor00}W. Wernsdorfer, R. Sessoli, A. Caneschi, D. Gatteschi, A.
Cornia, and D. Mailly, J. Appl. Phys. \textbf{87}, 5481 (2000).

\bibitem {Wernsdor01}W. Wernsdorfer, Advances in Chemical Physics,
\textbf{118}, 99(2001).

\bibitem {Wernsdor99b}W. Wernsdorfer, T. Ohm, C. Sangregorio, R. Sessoli, D.
Mailly, and C. Paulsen, Phys. Rev. Lett. \textbf{82}, 3903 (1999).

\bibitem {Wernsdor99c}W. Wernsdorfer, R. Sessoli, and D. Gatteschi, Europhys.
Lett. \textbf{47}, 254 (1999).

\bibitem {Prokof98}N. V. Prokof'ev and P.C.E. Stamp, Phys. Rev. Lett.
\textbf{80}, 5794 (1998).

\bibitem {Barbara99}B. Barbara, L. Thomas, F. Lointi, I. Chiorescu, and A.
Sulpice, J. Magn. Magn. Mater. \textbf{200}, 167 (1999).

\bibitem {Ohm98}T. Ohm, C. Sangregori, and C. Paulsen, Eur. Phys. J. B
\textbf{6}, 195(1998); J. Low Temp. Phys. \textbf{113}, 1141 (1998).

\bibitem {Cuccoli99}A. Cuccoli, A. Fort, A. Rettori, E. Adam, and J. Villain,
Eur. Phys. J. B \textbf{12}, 39 (1999).

\bibitem {Wernsdor00a}W. Wernsdorfer, A. Caneschi, R. Sessoli, D. Gatteschi,
A. Cornia, V. Villar, and C. Paulsen, Phys. Rev. Lett. \textbf{84}, 2965 (2000).

\bibitem {Chund00}E. M. Chudnovsky, Phys. Rev. Lett. \textbf{84}, 5676 (2000);
N. V. Prokof'ev and P.C.E. Stamp, \textit{ibid, }\textbf{84}, 5677 (2000) .

\bibitem {Liang00}J. Q. Liang , H. J. W. M\"{u}ller-Kirsten , D.K. Park, F.
\ C. Pu, Phys. Rev. B\textbf{\ 61}, 8856 (2000); Y. H. Jin, Y. H. Nie, J. Q.
Liang, Z. D. Chen, W. F. Xie, and F. C. Pu, Phys. Rev. B\textbf{\ 62}, 3316 (2000).

\bibitem {Chund01}E. M. Chudnovsky and D.A. Garanin, Phys. Rev. Lett.
\textbf{87}, 187203 (2001); D.A. Garanin and E. M. Chudnovsky, Phys. Rev. B
\textbf{65}, 094423 (2002).

\bibitem {LZ}L. Landau, Phys. Z. Sowjetunion, \textbf{2}, 46(1932); C. Zener,
Proc. R. Soc. London, Ser. A\textbf{137}, 696(1932).

\bibitem {Miyashita95}S. Miyashita, J. Phys. Soc. Jpn. \textbf{64}, 3207
(1995); \textit{ibid }\textbf{65}, 2734 (1996).

\bibitem {Garg93}A. Garg, Europhys. Lett. \textbf{22}, 205 (1993).

\bibitem {Chen02}Z. D. Chen, Phys. Rev. B \textbf{65}, 085313 (2002).

\bibitem {Palmer84}R.G. Palmer, D.L. Stein, E. Abrahams, and P.W. Anderson,
Phys. Rev. Lett. \textbf{53}, 958 (1984).

\bibitem {Erhart94}P. Erhart, A.M. Portis, B. Senning, and F. Waldner, J.
Phys. Condens. Matter \textbf{6}, 2881(1994); \ P. Erhart, B. Senning, and F
Waldner, J. Phys. Condens. Matter \textbf{6}, 2893(1994).

\bibitem {Chund88}E. M. Chudnovsky and L. Gunther, Phys. Rev. Lett.
\textbf{60}, 661(1988).

\bibitem {Chen}Z. D. Chen, J.-Q. Liang, S. Q. Shen, Phys. Rev. B, \textbf{66}, 092401(2002)

\bibitem {note1}Numerical result shows that the resulted relaxation curve is
not sensitive to the distribution center. In the present paper, we present
results for $h_{0}=0$ or $h_{0}=\sigma/4$ according to experimental
result.\cite{Ohm98}

\bibitem {Mertes01}K. M. Mertes, Y. Suzuki, M. P. Sarachik, Y. Paltiel, H.
Shtrikman, and E. Zeldov, E. Rumberger and D. N. Hendrickson, and G. Christou,
Phys. Rev. Lett. \textbf{87}, 227205 (2001).
\end{thebibliography}
\end{document}